\documentstyle[editedvolume,numreferences]{crckapb10}
\newcommand{\stt}{\small\tt}

\begin{opening}
\title{CURRENT STATUS~ OF ~RADIO SOURCE DATABASES}
\subtitle{}

\author{H. ANDERNACH}
\institute{Dpto.\,de Astronom\'\i a, IFUG, Apdo.\,Postal 144, Guanajuato, Mexico}
\author{S. TRUSHKIN}
\institute{Special Astrophysical Observatory, Nizhnij Arkhyz, 357147 Russia}

\end{opening}

\runningtitle{RADIO SOURCE DATABASES}

\begin{document}
\ifx\href\undefined\else\errmessage{Don't use hypertex!}\fi
\vspace*{-7.5cm}
\begin{footnotesize}\baselineskip 10 pt\noindent
Proc. {\it Observational Cosmology with the new Radio Surveys}, Tenerife, 
Spain, Jan.\,13--15, 1997 \\
eds.~~M.\,Bremer, N.\,Jackson \& I.\,P\'erez-Fournon, Kluwer Acad.\,Press, 
in press
\end{footnotesize}
\vspace*{6.4cm}

\section{The Electronic Source Catalogue Collection}

The realization \cite{And90} that astronomical data centers and databases 
showed a notable lack in published information on sources of radio emission,
motivated the first author in 1989 to start collecting and restoring electronic
versions of new and old source lists. By late 1993 more than 100
tables had been archived, and 67 of them, with $\sim$520,000 records
had been documented and made searchable via the ``Einstein On-line Service''
(EOLS) at CfA \cite{Har95}. In 1994, when EOLS lost financial support
and catalogue integration stagnated, a group around the second author
had independently started building the software tools to search and 
cross-identify radio sources from major catalogues. The latter evolved into
``CATS'' \cite{Ver95} maintained at Special Astrophysical Observatory 
(SAO, Russia), and is described in Section~2.
The table collection of the first author is being folded into CATS.

By Febr.\,1997 this largest existing set of electronic radio source lists
contains tables from over 400 different publications with over 
2$\times$10$^6$ records.
Just over 100 lists were collected in 1996 only. The current growth rate
is $\sim$12 items per month. Over 25\% of the 400 lists were prepared using 
a scanner, OCR software and strict proof-reading procedures.  
Most of the latter (but not all) were older lists, and some of them even
had to be typed by hand.
All 5C and Penticton "P"-surveys are now available and we
are close to completing the restoration of all published WSRT survey lists.
Virtually all source lists received from the authors had to be actively
requested by us. However, an increasing fraction of currently published
source lists (mostly smaller ones) can be found on preprint servers like
those at SISSA or LANL. Our experience is that almost half of all collected source
lists show some kind of problem either in the formatting, nomenclature or
other, requiring a subsequent interaction with the author for
clarification.  While only a minor fraction of the collected source lists has
been accessible via public databases before, we are providing public ftp 
access to an increasing number of items in our catalogue collection via 
CATS (see below). This is slowed down only by the necessity to compose
on-line documentation, usually not provided by the authors.

\section{CATS -- Astrophysical CATalogs support System}

CATS was developed by O.\,Verkhodanov, S.\,Trushkin \& V.\,Chernenkov
at SAO, primarily to support RATAN-600 radio observations.
CATS runs under LINUX and can process requests on the basis of
various net protocols and via email. It is accessible under URL~
{\tt http://www.ratan.sao.ru/$\sim$cats/}

About 70 well-known radio source lists with about 1.3$\times$10$^6$ 
records are now available via ftp from CATS.
Almost all catalogues with more than 2000 records of both Galactic and 
extragalactic origin were included in CATS.
Hypertext documentation of the database as well as adequate descriptions
of astrophysical catalogues are being prepared. These are based on over
150 original scientific publications and occupy 1.5\,Mb of CATS' 
current total size of $\sim$350 Mb.

Presently 16 of the larger tables may be searched simultaneously for objects
in rectangular boxes of coordinates or queried by additional parameters 
such as spectral index, flux density range, angular size, etc.

We use C routines ({\stt c\_sel}, {\stt c\_match}) to translate between
the content (RA,\,DEC, flux, etc.) and exact location of ASCII table columns.
This allows rapid folding of new catalogues into the search procedures.
We hope to have most CATS tables ready for searching soon.
The options {\stt select} and {\stt match}
allow to retrieve sources in boxes or circles around positions in either 
equatorial (B or J) or Galactic coordinates (cf. \cite{ADBIS96}).
The latest version of the NVSS catalogue \cite{Cond97} is available 
from CATS with correction routines rewritten in C and with a `CATS-like' 
interface.

CATS is being used for various astrophysical research projects. For example,
a cross-identification between infrared sources from the IRAS-PSC and
TXS radio sources at 365\,MHz yielded more than 1000 matches \cite{Tru96}. 
Most of these sources are likely to be of Galactic origin, because of their
clear concentration to the Galactic plane.  Cross-identification of
different catalogues are certainly an essential tool to investigate cosmic
objects (see \cite{Ver97} for another example).

The CATS authors consider creating some interpretation level CGI-programs
and various useful graphical presentations (as Java scripts) of results
of searches by user-defined parameters, e.g.\ radio spectra of selected 
cross-identified sources may be plotted in order to select sources
with peculiar spectra or to find variable objects. The latter are of
special interest for many cosmological questions.\\[-1ex]

H.A. acknowledges a travel grant from the conference organizers.
CATS is supported by RFBR grant N 96-07-89075.\\[-4.ex]

{}

\end{document}